\documentclass{article}
\usepackage{fixltx2e}
\usepackage{fix-cm}
\usepackage{amssymb}
\usepackage{amsmath}
\usepackage{graphicx}
\usepackage{makeidx}
\usepackage{multicol}
\usepackage{lipsum}
\usepackage{pinlabel}
\usepackage{enumerate}

\usepackage{color}

\newcommand{\BSS}{\mathrm{BSS}}

\newcommand{\SOL}{\mathrm{SOL}}
\newcommand{\SOLEVAL}{\mathrm{SOLEVAL}}

\newcommand{\N}{{\mathbb N}}

\newcommand{\FC}{\mathrm{FC}}
\newcommand{\FE}{\mathrm{FE}}

\newcommand{\Z}{{\mathbb Z}}
\newcommand{\R}{{\mathbb R}}
\newcommand{\C}{{\mathbb C}}

\newcommand{\bP}{{\mathbf P}}

\newcommand{\bNP}{{\mathbf{NP}}}
\newcommand{\bFPT}{{\mathbf{FPT}}}
\newcommand{\VNP}{{\mathbf{VNP}}}
\newcommand{\PSpace}{{\mathbf{PSpace}}}

\newtheorem{thm}{Theorem}
\newtheorem{theorem}[thm]{Theorem}

\newtheorem{definition}[thm]{Definition}

\newtheorem{examples}[thm]{Examples}
\newtheorem{prop}[thm]{Proposition}
\newtheorem{proposition}[thm]{Proposition}
\newtheorem{problem}{Problem}

\newif\ifsl 
\slfalse

\ifsl
\usepackage{showlabels}
\showlabels{bibitem}
\showlabels{index}
\fi

\newif\ifskip
\skiptrue 
\newif\ifmargin
\marginfalse
\newcommand{\JVW}{\mathrm{JVW}}
\newcommand{\NP}{\mathrm{\mathbf{NP}}}
\newcommand{\VP}{\mathrm{\mathbf{VP}}}
\newcommand{\sP}{\mathrm{\mathbf{\sharp P}}}
\newcommand{\FPT}{\mathrm{\mathbf{FPT}}}
\renewcommand{\P}{\mathrm{\mathbf{P}}}
\newcommand{\FP}{\mathrm{\mathbf{FP}}}
\newcommand{\sETH}{\mathrm{\mathbf{\sharp ETH}}}
\newcommand{\sWone}{\mathrm{\mathbf{\sharp W[1]}}}

\begin{document}

\title{The exact complexity of the Tutte polynomial}
\author{Tomer Kotek\\
Institute for Information Systems, Vienna University of Technology\\
Vienna 1040, Austria\\
\texttt{kotek@forsyte.tuwien.ac.at}\\
and \\
Johann A. Makowsky\\
Faculty of Computer Science, Israel Institute of Technology\\
Haifa 32000, Israel\\
\texttt{janos@cs.technion.ac.il}}
\maketitle
Longer version of Chapter 25 of the
{\em CRC Handbook
on the Tutte polynomial and related topics},
edited by 
J. Ellis-Monaghan
and 
I. Moffatt 

The publication of the Handbook has been delayed and is now scheduled for the first quarter of 2020.
The Chapter numbers in the Handbook may have changed. This also may affected the crossreferences
to be found in the posted paper.

In the version to be published in the Handbook the Sections 5 and 6 are shortened
and made into a single section.

\vspace{1cm}
\begin{center}
\includegraphics{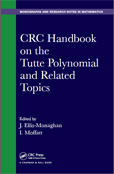}
\end{center}
\newpage
\tableofcontents
\newpage

\ifmargin
\marginpar{\color{red} File: NEW-synopsis}
\fi 
\newpage

\section{Synopsis}
In this chapter we explore the complexity of exactly computing the Tutte
polynomial and its evaluations for graphs 
and matroids in various models of
computation.
The complexity of approximating the Tutte polynomial is discussed in
{\color{red} Chapter 26, Bordewich}.
\begin{itemize}
\item
The Turing complexity of evaluating the Tutte polynomial exactly and the resulting
dichotomy theorem.
\item
Special attention is given to the case of
planar and bipartite planar graphs. 
\item
The impact of various notions of graph width (tree-width, clique-width,
    branch-width) on computing the Tutte polynomial
\item
The role of encoding matroids as inputs for computing the Tutte polyno-
    mial: Turing complexity for succinct presentations of matroids vs matroid
    oracles.
\item
The complexity in algebraic models of computation: Valiant's uniform fam-
    ilies of algebraic circuits versus the computational model of Blum-Shub-Smale
    (BSS).
\item
Open problems.
\end{itemize}

\ifmargin
\marginpar{\color{red} File: NEW-main}
\fi 
\ifmargin
\marginpar{\color{red} file: FINAL-intro.tex}
\else\fi 
\section{Introduction}
\label{KM:se:intro}

The bivariate Tutte polynomial $T(G;x,y)$ for graphs and matroids, introduced in 1954 by W.T. Tutte, arose from attempts to generalize the chromatic polynomial
introduced
by G. Birkhoff in 1912,
\cite{ar:Birkhoff1912,ar:Tutte54}.
The Tutte polynomial lived a life in the shadows of mainstream mathematics
until it entered center stage via the discovery in
\cite{jaeger1988tutte}
that the Jones polynomial
for knots and links was an incarnation
of the Tutte polynomial for alternating link diagrams. 
For the connection between the Tutte polynomial and knot theory, see 
{\color{red} Chapter 13 (S. Huggett)}.
\ifmargin
\marginpar{\color{red} Chapters cross reference}
\else \fi 
Since then
numerous papers have investigated amazing properties of the Tutte polynomial and its generalization
to signed graphs and more generally to edge colored graphs, see 
{\color{red}
Chapter 1 (J. Ellis-Monaghan and I. Moffatt),
Chapter 18 (L. Traldi), and for its history, 
Chapter 29 (G. Farr).
}
\ifmargin
\marginpar{\color{red} Chapters cross reference}
\else \fi 

When evaluating the Tutte polynomial we look at problem of computing the exact value of $T(G; x_0, y_0)$ where
the input is a graph $G$ and two elements $x_0, y_0 \in \C$.
One could, alternatively, also ask for the list of all coefficients of $T(G; x_0, y_0)$,
or, more modestly, for the sign of $T(G; x_0, y_0)$, when $x_0, y_0 \in \R$.
If we can evaluate efficiently, we can use interpolation to compute the coefficients, see \cite{ar:JVW90}
for a detailed discussion. Surprisingly, M. Jerrum and L. Goldberg showed in \cite{goldberg2014complexity}
that even computing the sign of $T(G; x_0, y_0)$ can be very hard.
In this chapter we 
concentrate on the complexity of exactly evaluating the Tutte polynomial and its relatives
as a function of the order of the graph $G$.

Approximate computation of the Tutte polynomial is treated in
{\color{red} Chapter 26, Bordewich}.
The landmark paper initiating the study of the complexity of $T(G;x,y)$ is
\cite{ar:JVW90}.
This paper sets the paradigm for further investigations.
It studies the complexity of the problem, 
given a graph $G$ as input, 
to compute the value of all the coefficients
of the polynomial $T(G;x,y)$.
The question is stated and answered in the 
Turing \index{Turing}\index{computation model!Turing}
model of computation and the complexity class suitable to capture
the most difficult cases is $\sharp\bf{P}$, which was introduced by Valiant in \cite{valiant1979complexity}.
In the course of the chapter it is observed that
the results can be extended to any subfield of $\C$, 
the elements of which can be represented is finite binary strings.
The main result of \cite{ar:JVW90} states that for almost all pairs $(x_0,y_0) \in \C$ the problem is
$\sharp\bf{P}$-complete, and the 
set for which it is not $\sharp\bf{P}$-complete
(the exception set)
is given explicitly as a semialgebraic set of lower dimension.
We shall discuss this result in detail in Section \ref{KM:se:jvw}.

\ifmargin
\marginpar{\color{red} file: FINAL-background.tex}
\else\fi 
\section{Background I} 
\label{KM:se:background}
In this section we collect some background material needed for discussing the complexity of the Tutte polynomial
for graphs. Further background for the case of matroids follows in Section \ref{KM:se:matroids}.
\subsection{Complexity classes}
\index{complexity}
Background on complexity in the Turing model of computation can be found in 
\cite{bk:johnson,bk:papadimitriou94,sip97,goldreich2008computational,bk:AB2009}, and
for Valiant style algebraic complexity in \cite{bk:Buergisser00}.
For complexity over the reals and complex numbers we refer to \cite{bk:BCSS}, and
for quantum computing we refer to \cite{gruska1999quantum,nielsen2010quantum}.
For parameterized complexity we refer the reader to 
\cite{bk:FlumGrohe2006,downey2012parameterized,downey2013fundamentals,cygan2015parameterized,fomin2010exact}.

\index{complexity!decision problem}
Informally, a {\em decision problem} is a computational problem for which the answer is either 
'yes' or 'no'; e.g., given a graph $G$, is there a proper $r$-coloring of it?
\index{complexity!counting problem}
A {\em counting problem} is a computational problem for which the answer is the number of satisfying configurations;
e.g., given a graph $G$, how many proper $r$-colorings does it have? 
\index{complexity!polynomial time}
\index{complexity!tractable}
\index{complexity!P}
\index{complexity!FP}
We denote by $\P$, respectively $\FP$, the set of decision problems (counting problems) 
which can be solved in time polynomial in the size of the input.
Problems which belong to $\P$ or to $\FP$ are called {\em tractable}, and are considered to be efficiently computable. 

\index{complexity!NP}
We denote by $\NP$ the set of 
decision problems for which it is possible to {\em verify} whether a given configuration 
is correct in polynomial time in the size of the input. 
\ifsl
\index{complexity!sharp P}
\else
\index{complexity!sharp P@$\sharp \mathrm{P}$ (sharp P)}
\fi
We denote by $\sP$ (pronounced \emph{number P}) the set of counting problems which count 
configurations whose correctness can be verified in polynomial time.
For example the problem of deciding whether a graph is $r$-colorable is in $\NP$, 
and  computing the number of $r$-colorings is in $\sP$. 
Whether $\P = \NP$ and whether $\FP = \sP$ are 
famous and notoriously difficult open questions in theoretical computer science. 
\index{complexity!NP-hard}
\ifsl 
\index{complexity!sharp P-hard}
\else
\index{complexity!sharp P-hard@$\sharp \mathrm{P}$-hard}
\fi
A problem $X$ is {\em $\NP$-hard} ({\em $\sP$-hard}) if using it as an oracle allows one to solve any problem
in $\NP$, respectively in $\sP$, in polynomial time.
Problems which are $\NP$-hard or $\sP$-hard are provably the hardest problems in $\NP$, respectively $\sP$,
and are are considered {\em intractable}.
\index{complexity!intractable}
They are not considered  effectively computable, 
unless $\NP= \P$, respectively $\sP=\FP$. 

So far the complexity of computational problems was measured with respect to one parameter of the input only: the size of the input
(e.g.  the number of vertices or edges of the input graph). 
\index{complexity!parameterized}
{\em Parameterized complexity} measures the complexity of problems in terms
of additional parameters of the graph. A problem is {\em fixed-parameter tractable with respect to a parameter $k$} 
\index{complexity!fixed-parameter tractable}
if 
there is a computable function $f$, a polynomial $p$, and 
an algorithm solving the problem in time $f(k)\cdot p(n)$, where $n$ is the size of the problem,
and the function $f(k)$ does only depend on $k$. 
However, $f$ may grow arbitrarily in $k$, and indeed is  often exponential in $k$. 
The set of fixed-parameter tractable decision problems is denoted by $\FPT$. 
\index{complexity!FPT}

\subsection{Structural graph parameters}

\index{width|(}
\index{tree-width}
\index{clique-width}
\index{width!tree-width}
\index{width!clique-width}
Tree-width and clique-width are graph parameters which measure the compositionality of a graph. Tree-width is usually
defined as the minimum width of a certain map of a graph into a tree called a {\em tree-decomposition} 
In contrast, graphs of clique-width $k$ are defined inductively. 
For uniformity of presentation, we also give an inductive definition  of graphs of tree-width $k$. 

The various notions of 
graph width have had a large impact on the study of efficient algorithms for graph problems, 
\cite{ar:HlinenyOumSeeseGottlob08}. 
Due to the inductive definitions
of graphs of bounded tree-width and clique-width, problems which are $\NP$-hard or $\sP$-hard 
(on general graphs) are often fixed parameter tractable
with respect to tree-width or clique-width \cite{courcelle1990monadic,courcelle2001fixed}. 
Some of the techniques used to compute the Tutte polynomial on graphs of bounded tree-width
may also be found in 
{\color{red}Chapter 26 (C. Merino)}.
\ifmargin
\marginpar{\color{red} Chapter cross reference}
\else\fi 

\ifsl
\index{k-graph}
\else
\index{k-graph@$k$-graph}
\fi
Let $[k] =\{1,\ldots,k\}$. A {\em $k$-graph} is a 
graph $G=(V,E)$ together with a partition $\bar{R}=(R_1,\ldots,R_k)$ of $V$. 
The sets $R_i$ are called {\em labels} of $G$ and $\bar{R}$ is a {\em labeling}. The classes $TW(k)$ and $CW(k)$  of $k$-graphs are defined inductively below. A graph has {\em tree-width (clique-width)  $k$}
if $k$ is the minimal value such that there is a labeling $\bar{R}$ of $G$ for which $(G,\bar{R})\in TW(k)$, 
respectively in $(G,\bar{R})\in CW(k)$. 

\begin{definition}[Tree-width]
\index{tree-width}
The class $TW(k)$ of graphs of tree-width at most $k$ is defined inductively as follows:
\begin{itemize}
 \item[--] All $(k+1)$-graphs of order at most $k+1$ belong to $TW(k)$.
 \item[--] $TW(k)$ is closed under the following operations:
 \begin{enumerate}[(i)]
 \item Disjoint union $\sqcup$;
 \item Renaming of labels $\rho_{i\to j}$: all vertices in $R_i$ are moved to $R_j$;
 \item Fusion $\mathit{fuse}_i$: all vertices in $R_i$ are contracted into a single vertex.  
 \end{enumerate}
\item[--] $tw(G)$, the tree-width of $G$, denotes the smallest $k$ such that $G \in TW(k)$.
\end{itemize}
\end{definition}
All trees have tree-width $1$ and all cycles have tree-width $2$. A clique of size $k$ has tree-width $k-1$. 

\begin{definition}[Clique-width]
\index{clique-width}
The class $CW(k)$ of graphs of clique-width at most $k$ is defined inductively as follows:
\begin{itemize}
 \item[--] All $k$-graphs of order $1$ belong to $CW(k)$.
 \item[--] $CW(k)$ is closed under the following operations:
 \begin{enumerate}[(i)]
 \item Disjoint union $\sqcup$;
 \item Renaming of labels $\rho_{i\to j}$;
 \item Edge addition: $\eta_{i,j}$: all possible undirected edges are added between $R_i$ and $R_j$. 
 \end{enumerate}
\item[--] $cw(G)$, the clique-width of $G$, denotes the smallest $k$ such that $G \in CW(k)$.
\end{itemize}
\end{definition}

Every graph of tree-width at most $k$ has clique-width at most $2^{k+1}+1$, cf. \cite{ar:CourcelleOlariu2000},
while there are classes of graphs of bounded clique-width which have unbounded tree-width, 
such as cliques (clique-width $1$) or complete bipartite graphs (clique-width $2$).
{\em Logarithmic clique-width} is defined similarly to clique-width, 
with the exception that $R_1,\ldots,R_k$ are no longer required to be 
disjoint.
Every graph of tree-width at most $k$ has logarithmic clique-width at most $k+2$.
A graph $G=(V,E)$ of tree-width $k$ has at most $k|V|$ edges, 
while a graph of clique-width $k$ can have the maximal number of edges of a 
loop-free undirected graph $\binom{|V|}{2}$. 

\ifsl
\index{k-expression}
\else
\index{k-expression@$k$-expression}
\fi
A $TW(k)$-expression 
is a term $t$ consisting of base elements and 
operations which witnesses that a $k$-graph belongs to $TW(k)$. ($CW(k)$). 
A $CW(k)$-expression  is defined correspondingly.
Computing the exact tree-width or the exact clique-width of a graph is $\NP$-hard. 
However, 
computing a $TW(k)$-expression is fixed parameter tractable in $k$
due to the fixed parameter tractability of computing a tree decomposition \cite{ar:Bodlaender1996}. 
For clique-width,  the computation of an exponential upper bound $k'$ on the clique-width $k$ of a graph
and a $CW(k')$-expression for the graph
is fixed parameter tractable \cite{ar:OumSeymour06}.
\index{width|)}
Other notions of widths of graphs are studied in the literature. A survey is given in
\cite{ar:HlinenyOumSeeseGottlob08}. Among these we have 
$\mathrm{bw}(G)$, the {\em branch-width} of $G$, which is related to tree-width, and 
$\mathrm{rw}(G)$, the {\em rank-width} of $G$, which is related to clique-width. 
More precisely,
\begin{theorem}
\begin{enumerate}[(i)]
\item
$\mathrm{bw}(G) \leq \mathrm{tw}(G) +1 \leq \lfloor \frac{3}{2}\mathrm{bw}(G)\rfloor$,
\cite{robertson1991graph}.
\item
$\mathrm{rw}(G) \leq \mathrm{cw}(G) \leq 2^{1+\mathrm{rw}(G)}-1$,
\cite{ar:OumSeymour06}.
\end{enumerate}
\end{theorem}
Among these notions of width only branch-width generalizes to matroids, 
see Section \ref{KM:se:matroid-width}.

\index{computation model!Turing|(}
\ifmargin
\marginpar{\color{red} file: ARXIV-graphs.tex}
\else\fi 
\section{The exact complexity of the Tutte polynomial on graphs}
\label{KM:se:graphs}

\subsection{The main paradigm: A dichotomy theorem}
\label{KM:se:jvw} 
\label{se:4-1}

\index{dichotomy theorem!graphs}
The natural basic computational task associated with the Tutte polynomial is to compute, for a given graph $G$,
the table of coefficients of $T(G;x,y)$. 
However, this task is $\sP$-hard, since the number $(-1)^{r(E)} 3^{-k(G)} T(G;-2,0)$ of proper $3$-colorings of $G$ is $\sP$-hard, 
and the prefactor $(-1)^{r(E)} 3^{-k(G)}$ is polynomial-time computable. 
In particular, the evaluation $(-2,0)$ of the Tutte polynomial is $\sP$-hard. 

\index{dichotomy theorem!Tutte polynomial}
F. Jaeger, D. Vertigan and D. Welsh began the study of the complexity of the Tutte polynomial in their classic paper \cite{ar:JVW90}. They studied the problem of computing the Tutte polynomial
on input a graph $G$ at a fixed point $(a,b)\in \mathbb{C}^2$ in the complex plane.  The main result of \cite{ar:JVW90} is a {\em dichotomy theorem}, a theorem which classifies the complexity of evaluations $(a,b)$
of the Tutte polynomial into {\em tractable}  or {\em $\sP$-hard}. 
Technically, the complexity is measured in an extension field of the rational numbers
containing $a$ and $b$.

Let $H_1$ be the hyperbola 
\[ H_1 = \{(x,y) \in \C^2: ((x-1)(y-1)=1) \}\,.\]
\begin{thm}[F. Jaeger, D. Vertigan and D. Welsh \cite{ar:JVW90}] 
\label{CRC:JVW-theorem}
\ \\
Let $G$ be a graph, and
let $H$ be the union of $H_1$ with
$$\{(1,1),(-1,-1),(0,-1),(-1,0),(i,-i),(-i,i),(j,j^{2}),(j^{2},j)\},$$
where $j=e^{2\pi i/3}$. For every $(a,b)\in\mathbb{C}^2$:
\begin{enumerate}
\item If $(a,b)\not\in H$, then $T(G;a,b)$ is $\sP$-hard.
\item If $(a,b)\in H$, then $T(G;a,b)$ is computable in polynomial-time. 
\end{enumerate}
\end{thm}

\begin{figure}
 \centering
    \includegraphics[width=0.4\textwidth]{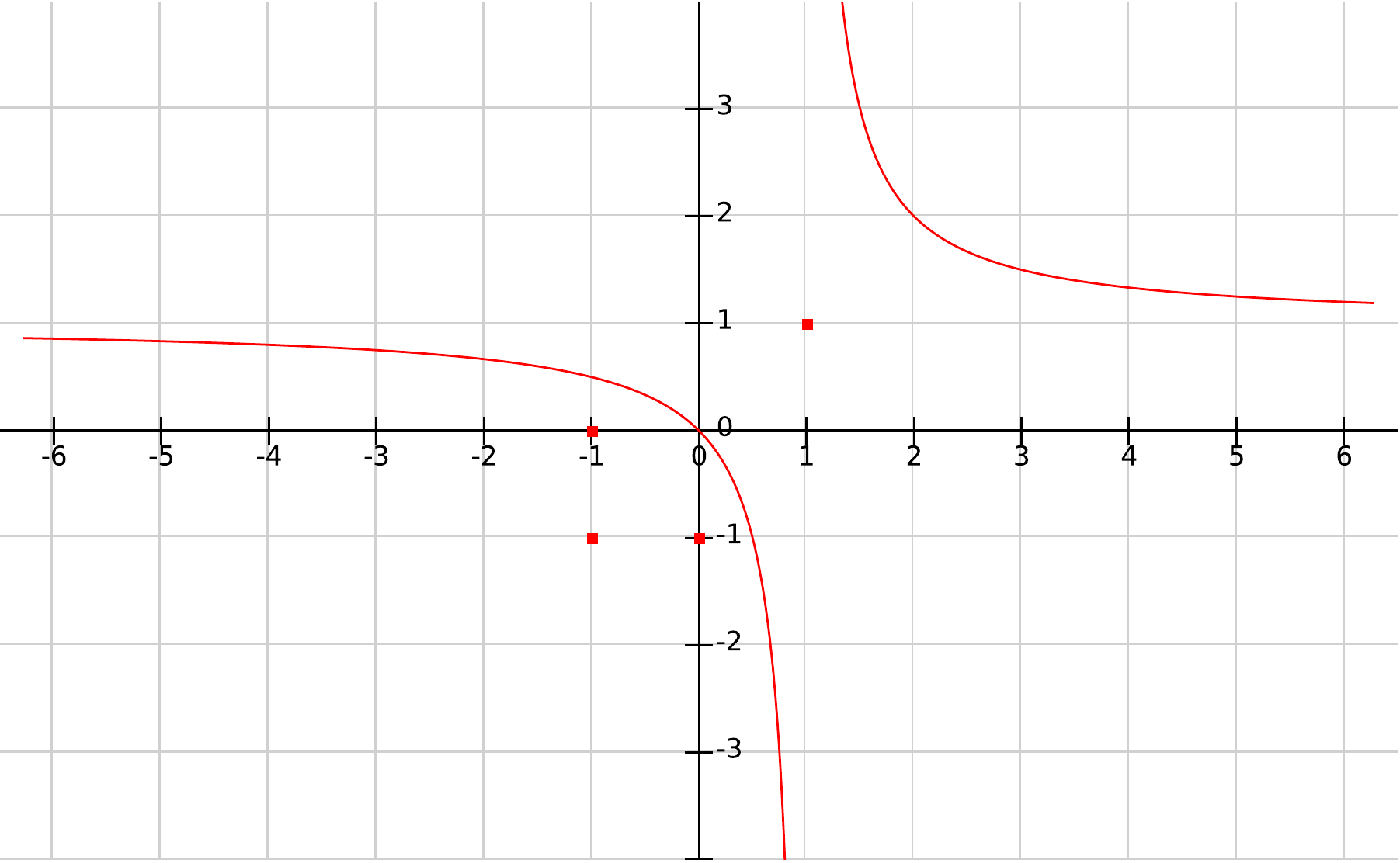}\ \ \ \ 
 \caption{\label{KM:fig:JVW} $T(G;x,y)$ is polynomial-time computable on the hyperbola $(x-1)(y-1)=1$ and the points $(1,1),(-1,-1),(0,-1),(-1,0)$. 
 Otherwise it is $\sP$-hard. 
 }
 
\end{figure}

See Figure \ref{KM:fig:JVW} for a plot of the $\R^2$-part of the Tutte  plane. 

\subsection{Planar and bipartite planar graphs}
\label{se:4-2}

The method of Kasteleyn \cite{ar:Kasteleyn61} for tractable computation of the Ising partition
function on planar graphs carries over to the Tutte polynomial.

\begin{thm}[P. Kastelyn, \cite{ar:Kasteleyn61}]
\label{th:Kastelyn}
For planar graphs
evaluating  the Tutte polynomial 
on the hyperbola \[H_{2}\equiv((x-1)(y-1)=2)\,,\]
is in $\bP$.
Otherwise, points which are $\sP$-hard on general graphs remain $\sP$-hard even on bipartite planar graphs. 
\end{thm}

\index{dichotomy theorem!bipartite planar graphs}

Jaeger and Welsh studied the complexity of evaluating the Tutte polynomial on the class of 
graphs which are both bipartite and planar. 

\begin{thm}
[D. Vertigan and D. Welsh \cite{ar:VW92}, \cite{ar:Vertigan05}]
\label{th:VW}
\ \\
Let $H_{\mathit{b-p}}$ be the union of $H_{1}$, $H_2$, 
and 
$\{(1,1),(-1,-1),(j,j),(j,j)\}$.
For every $(a,b)\in\mathbb{C}^2$:
\begin{enumerate}
\item If $(a,b)\not\in H_{b-p}$, then $T(G;a,b)$ is $\sP$-hard for  bipartite planar graphs. 
\item If $(a,b)\in H_{b-p}$, then $T(G;a,b)$ is computable in polynomial-time for bipartite planar graphs. 
\end{enumerate}
\end{thm}


\begin{figure}
 \centering
    \includegraphics[width=0.4\textwidth]{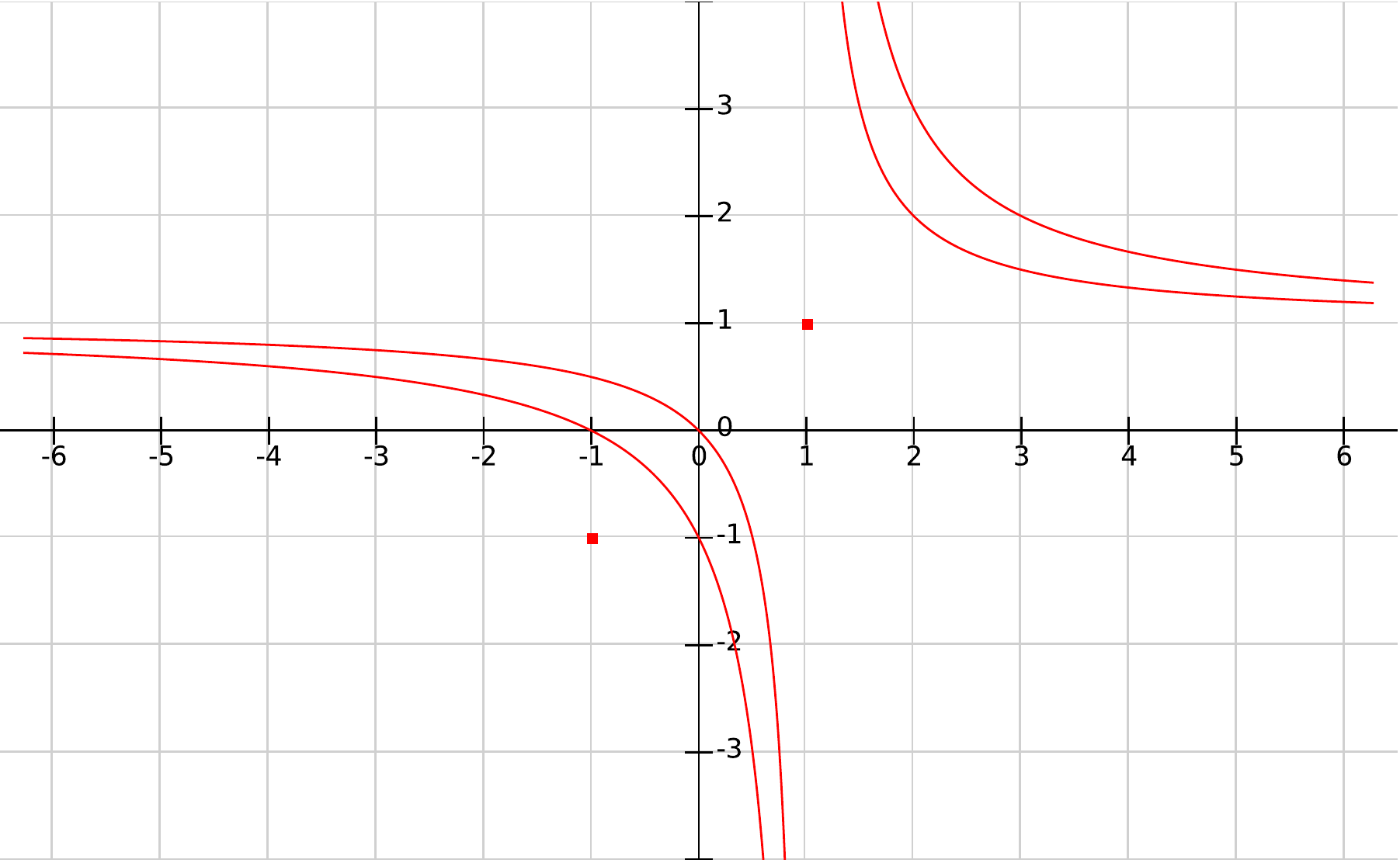}\ \ \ \ 
 \caption{\label{KM:fig:VW} $T(G;x,y)$ is polynomial-time computable on the hyperbolas $(x-1)(y-1)=\{1,2\}$ and the points $(1,1),(-1,-1)$
 for bipartite planar graphs. 
 Otherwise it is $\sP$-hard. 
 }
\end{figure}

See Figure \ref{KM:fig:VW} for a plot of the Tutte plane for bipartite planar graphs. 

\subsection{Graphs of bounded tree-width and clique-width}
\label{se:4-3}
The Tutte polynomial and variants of it are efficiently computable on graphs of bounded tree-width
or clique-width. 
\index{tree-width|(}
\index{width!tree-width|(}

J. Oxley and D. Welsh initiated the study of the complexity of evaluating the Tutte polynomial
on graph classes of bounded width, \cite{oxley1992tutte}.
However, their notion of width is more restricted than tree-width, but does include the series-parallel graphs,
which are of tree-width $2$.
The first general results on the computation of the Tutte polynomial on graphs of 
bounded tree-width were obtained independently by
A. Andrzejak \cite{ar:andrzejak98} and S. Noble \cite{ar:noble98}. They proved that for every 
fixed evaluation $(a,b)$ of $T(G;x,y)$, there an algorithm which computes $T(G;a,b)$ using a linear number of
arithmetic operations. 
\begin{thm}
[A. Andrzejak \cite{ar:andrzejak98}, S. Noble \cite{ar:noble98}] \label{KM:th:tutte-tw}
\label{th:A-N}
\ \\
$T(G;x,y)$ can be evaluated in linear time
in $|V|$ on graphs with bounded tree-width. In fact, $T(G;x,y)$
is fixed-parameter tractable with respect to tree-width.
\end{thm}

Analogues of Theorem \ref{KM:th:tutte-tw} for variants of the Tutte polynomial are given below. 
The coefficients of the Tutte polynomial and its variants can be computed in polynomial-time 
on graphs of bounded tree-width. 
In the case of $T(G;x,y)$, the list of coefficients can be computed in time $O(|V|^3)$ by evaluating $T(G;x,y)$
at sufficiently many points to interpolate it. 
There is a matching cubic lower bound for the computation of all the coefficients \cite{ar:noble98}. 

The result \cite{ar:andrzejak98} and \cite{ar:noble98}
has a wide generalization  based on logical techniques
which covers colored versions of the Tutte polynomial, the Jones polynomial in knot theory, 
and many other graph polynomials which can be defined in {\em Monadic Second Order logic}.

\index{normal function of the colored matroid}
T. Zaslavsky \cite{zaslavsky1982chromatic,ar:Zaslavsky1992} defined a generalization of the Tutte polynomial 
with an edge coloring function $c:E\to\lambda$
called the 
{\em  the signed Tutte polynomial for graphs and} 
{\em normal function of the colored matroid}\footnote{
We discuss the complexity of the Tutte polynomial on matroids in Section \ref{KM:se:tmatroids}. 
}
, respectively. This was rediscovered by
Bollob\'{a}s and Riordan \cite{bollobas-riordan1999tutte} as colored versions of the Tutte polynomial.
\index{colored Tutte polynomial}

\begin{thm}[I. Averbouch, B. Godlin and J. Makowsky \cite{ar:AGM10}, \cite{makowsky2005coloured}]
\ 
\begin{enumerate}[(i)]
\item
Evaluating the Tutte polynomial for signed graphs and normal function of a colored  graph 
is fixed-parameter tractable with respect to tree-width, \cite{ar:AGM10}.  
\item
Let $\Lambda$ be a finite set. 
Evaluating the colored Tutte polynomial 
$$
T_{\mathit{colored}}(G,c;x_\lambda,y_\lambda,X_{\lambda},Y_\lambda:\,\lambda\in \Lambda)
$$
for colored graphs $(G,c)$ with $c:E\to\Lambda$
is fixed-parameter tractable with respect to tree-width, 
\cite{makowsky2005coloured}.
\end{enumerate}
\end{thm}
Similar theorems stated for knot diagrams rather than graphs can be found in  \cite{makowsky2003parametrized}.
A detailed discussion of the signed and colored Tutte polynomials can be found in 
Chapter 17.1.
The Jones polynomial and other relatives of the Tutte polynomial
and their application to knot theory are discussed in
{\color{red}Chapter NN (X. YYY)}.
\ifmargin
\marginpar{\color{red} Chapter cross reference}
\else\fi 

The complexity of the algorithms given in  
\cite{makowsky2005coloured} and \cite{ar:AGM10} 
is asymptotically the same as in
\cite{ar:andrzejak98} and \cite{ar:noble98}. 
However,
the runtimes of the algorithms
given in
\cite{makowsky2005coloured} and \cite{ar:AGM10} 
work also for the 
signed or colored Tutte polynomials, but
involve very large constants due to the generality of 
the logical methods involved which make them impractical.
The algorithms given in
\cite{ar:andrzejak98} and \cite{ar:noble98} give workable upper bounds, but do not work
for the signed or colored Tutte polynomials.
A more efficient algorithm for the colored Tutte polynomial 
was given in \cite{ar:Traldi2006} based on \cite{ar:andrzejak98}.


\ifsl
\index{the weighted graph polynomial U}
\else
\index{the weighted graph polynomial@the weighted graph polynomial $U$}
\fi
Finally, Noble in \cite{ar:noble2009} extended his approach to a further generalization of the Tutte polynomial,
the weighted graph polynomial $U$.
He showed that 
evaluating the weighted graph polynomial $U(G;\bar{x},y)$ is also 
fixed-parameter tractable with respect to tree-width. 

\index{tree-width|)}
\index{width!tree-width|)}


On graphs of bounded clique-width
the Tutte polynomial can be computed in subexponential time.
However,
unlike the case of tree-width, $T(G;x,y)$ is unlikely to be fixed-parameter tractable with respect to clique-width.
\index{clique-width|(}
\index{width!clique-width|(}
\index{complexity!subexponential}

\begin{thm}
[O. Gim\'{e}nez, P. Hlin\v{e}n\'{y} and M. Noy \cite{gimenez2006computing}] 
\label{KM:th:CW-tutte}
\label{th:GHN}
\ \\
$T(G;x,y)$  can be computed in time $\exp\left(O(|V|^{1-1/(k+2)})\right)$
on graphs of clique-width at most $k$. 
\end{thm}

\begin{thm}
\ \\
{\bf (F. Fomin, P. Golovach, D. Lokoshtanov and S. Saurabh \cite{ar:FominGolovachLokshtanov10,fomin2010intractability})}
\ \\
Given a graph $G$ and $r\in\N$, deciding whether $\chi(G;r)\neq 0$
is not fixed-parameter tractable with respect
to clique-width 
(under the complexity-theoretic assumption $\FPT\not=\sWone$). 
In particular, this shows that the same applies to evaluating the chromatic polynomial.
\end{thm}

The chromatic polynomial, however, is polynomial-time computable if the clique-width is fixed:

\begin{thm}
\label{th:MRAG}
\ \\
{\bf (J. Makowsky, U. Rotics, I. Averbouch and B. Godlin \cite{makowsky2006computing})}
\ \\
The chromatic polynomial $\chi(G,\lambda)$ can be  computed
 in time $O(|V|^{f(k)})$
on graphs of clique-width at most $k$,
where $f(k)= O(2^k)$ is a function depending only on $k$.
\end{thm}

\index{clique-width|)}
\index{width!clique-width|)}

It is open whether Theorem \ref{th:GHN} can be improved to match the bound for the chromatic polynomial given in
Theorem \ref{th:MRAG}.

We summarize the results of Sections \ref{se:4-1}-\ref{se:4-3} in Table \ref{table-1}.

\small
\begin{center}
\begin{table}[htb]
\begin{tabular}{|l | l| l| l | l| c |}
\hline
\hline
&&&&& \\
Graph class & $\sharp \bP$-hard & subexponential & $\bFPT$ & $\bP$ & Theorem\\
&&&&&\\
\hline
\hline
\vspace{0.05cm}
&&&&& \\
All graphs & $\C^2 -H$ & $H$ & $H$ & $H$ & \ref{CRC:JVW-theorem}\\
\hline
\vspace{0.05cm}
&&&&& \\
planar & $\C^2 -H_2$ & $H_2$ & $H_2$ & $H_2$ & \ref{th:Kastelyn}\\
\hline
\vspace{0.05cm}
&&&&& \\
bipartite planar & $\C^2 -H_{b-p}$ & $H_{b-p}$ & $H_{b-p}$ & $H_{b-p}$ & \ref{th:VW}\\
\hline
\vspace{0.05cm}
&&&&& \\
$TW(k)$ & $\emptyset$ & $\C^2$ & $\C^2$ & $H$ & \ref{th:A-N}\\
\hline
\vspace{0.05cm}
&&&&& \\
$CW(k)$ & $\emptyset$ & $\C^2$ & $H$ & $H$ & \ref{th:GHN}\\
\hline
\end{tabular}
\caption{The complexity of evaluating the Tutte polynomial for various graph classes}
\label{table-1}
\end{table}
\end{center}

.

\subsection{Exact computation of the $T(G;x,y)$ on small graphs}

\index{explicit computation}
Computing the Tutte polynomial on large graphs is a major challenge. The deleting-contraction reduction formula
of the Tutte polynomial gives a naive algorithm for computing the list of coefficients of the Tutte polynomial
whose runtime is exponential in the number of edges of the graph. 
Sekine, Imai and Tani \cite{sekine1995computing} gave an algorithm computing the Tutte polynomial
of a planar graph of order $n$
whose runtime is  $O(2^{O(\sqrt{n})})$.
Bj\"{o}rklund et al. \cite{bjorklund2008computing} gave an algorithm for general graphs which is exponential in the number
of vertices and uses polynomial space. 
Haggard, Pearce and Royle \cite{haggard2010computing}
implemented an algorithm which 
exploits isomorphisms in the computation tree to allow more efficient computation of the Tutte polynomial
and used it disprove a conjecture by Welsh on the roots of the Tutte polynomial. 
See {\color{red} Chapter XX} 
\ifmargin
\marginpar{\color{red} Chapter cross reference}
\else
\fi 
for more details. 

\subsection{An exponential-time dichotomy theorem}
\index{dichotomy theorem!exponential time}
\index{complexity!subexponential time}
An algorithm has {\em subexponential running time} if it runs in time $2^{n^{o(1)}}$.
While there are subexponential algorithms for the Tutte polynomial on restricted classes of graphs
\cite{gimenez2006computing,sekine1995computing}, a subexponential algorithm for the class of all graphs
is unlikely to exist\footnote{Note that the problem 3-SAT of whether
propositional 3-CNF formulas are satisfiable is reducible via a linear-size reduction to the chromatic polynomial; 
A subexponential algorithm for 3-SAT is considered very unlikely. 
}.  
Dell et al. \cite{DellHMTW14}
gave a dichotomy theorem which matches the vertex-exponential algorithm of
 \cite{bjorklund2008computing} under the assumption that there is no subexponential algorithm 
 for the number of proper $3$-colorings. This assumption is called the {\em counting Exponential Time Hypothesis ($\sETH$)}.

\begin{thm}
\label{th:DHMTW}
\ \\
{\bf (H. Dell, T. Husfeldt, D. Marx, N. Taslaman, M. Wahlen \cite{DellHMTW14})} 
\\
Let $(a,b)\in\mathbb{Q}^{2}$. Under $\sETH$:
\begin{enumerate}[(i)]
\item $T(G;a,b)$ cannot be computed in  subexponential time in $|V|$ if $(a,b)\notin H_1$
and $b\notin\{-1,0,1\}$.
\item $T(G;a,b)$ cannot be computed in subexponential time in $|V|$ if $y=0$ and
$a\notin\{-1,0,1\}$ on simple graphs.
\item $T(G;a,b)$ cannot be computed in subexponential time in 
$(|E|/\log^{2}|E|))$ if $a=1$
and $b\not=1$ on simple graphs.
\item $T(G;a,b)$ cannot be computed in subexponential time in 
$(|E|/\log^{3}|E|))$ if $(a-1)(b-1)\not=\{0,1\}$
and $(a,b)\notin\{(-1,-1),(-1,0),(0,-1)\}$ on simple graphs. 
\end{enumerate}
The line $y=1$ is still open ($\sP$-hardness is the best known). 
\end{thm}
Using results of 
R. Curticapean \cite{Curticapean15} (iii) and (iv) can be improved by eliminating the
logarithmic factor.

\begin{figure}
 \centering
    \includegraphics[width=0.4\textwidth]{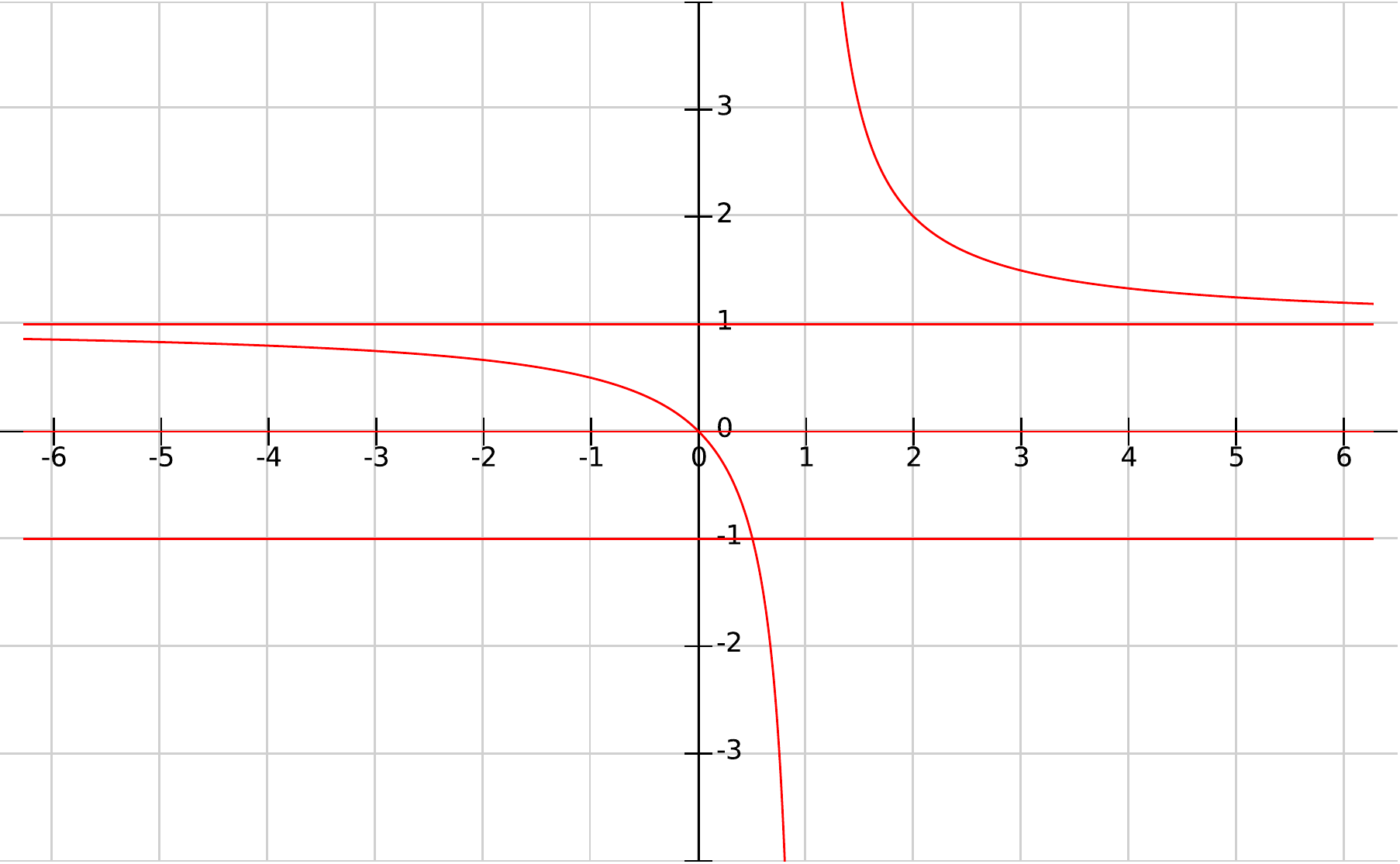}\ \ \ \ 
    \includegraphics[width=0.4\textwidth]{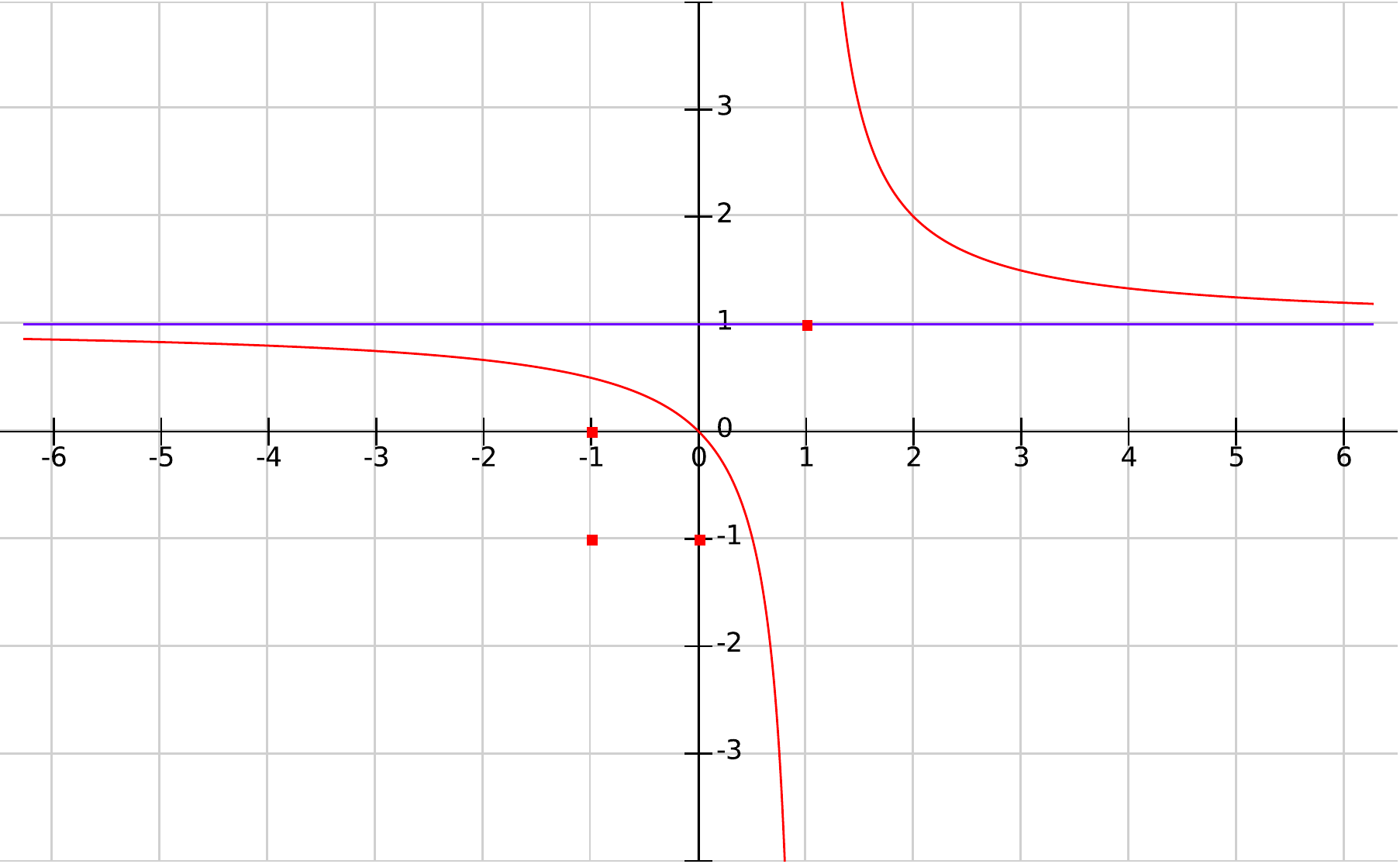}\\
    (a)
    \ \ \ \ \ \ \ \ \ \ \ \ \ \ \ \ \ \ \ \ \ \ \ \ \ \ \ \ \ \ \ \ \ \ \ \ \ \ \
    (b)
 \caption{\label{KM:fig:ETH} (a) Except for the hyperbola $(x-1)(y-1)=1$ and the lines $y\in \{0,-1,1\}$, $T(G;x,y)$ cannot be computed in subexponential vertex-time. (b)
 Except for the hyperbola $(x-1)(y-1)=1$, the line $y=1$ and the points $(-1,-1),(-1,0),(0,-1)$, $T(G;x,y)$ cannot be computed in time which is subexponential in
 $|E|/\log^3|E|$ on simple graphs. }
\end{figure}
See Figure \ref{KM:fig:ETH} for plots of subexponential lower bounds for $T(G;x,y)$. 

\ifmargin
\marginpar{\color{red} file: ARXIV-matroids.tex}
\else\fi 
\section{Background II}
\label{KM:se:matroids}
\label{KM:se:mbackground}
Matroids were introduced in 
\ifmargin
\marginpar{\color{red} Chapter cross reference}
\else\fi 
{\color{red}
Chapter 8 (J. Oxley) and further discussed in Chapter 11 (E. Gioan).
}
For convenience of the reader we repeat some of the basic definitions here.
A 
matroid 
\index{matroid}
is an ordered pair $(E, \mathcal{C})$ consisting of a finite set $E$, called the {\em ground set}
and a set $\mathcal{C}$ of subsets of $E$, called the {\em circuits} which satisfy the following conditions:
\begin{description}
\item[(C1)]
$\emptyset \not \in \mathcal{C}$
\item[(C2)]
If $C_1, C_2 \in \mathcal{C}$ and $C_1 \subseteq C_2$ then $C_1 =C_2$.
\item[(C3)]
If $C_1, C_2 \in \mathcal{C}, C_1 \neq C_2$ and $e \in C_1 \cap C_2$ then there is $C_3 \in \mathcal{C}$
such that $C_3 \subseteq (C_1 \cap C_2) - \{e\}$.
\end{description}
Matroids have many equivalent definitions, cf. \cite{bk:Welsh76,ar:HausmannKorte81} or 
{\color{red} Chapter 8, Section 4 (J. Oxley)}.
Instead of $\mathcal{C}$ one can consider 
$\mathcal{I}$, the {\em independent sets}, 
\index{matroid!independent sets}
$\mathcal{S}$, the {\em spanning sets},
\index{matroid!spanning sets}
$\mathcal{F}$, the {\em flats},
\index{matroid!flats}
$\mathcal{B}$, the {\em bases},
\index{matroid!bases}
$\mathcal{H}$, the {\em hyperplanes}.
\index{matroid!hyperplanes}
Other definitions use a {\em closure operator} on the subsets of $E$, 
\index{matroid!closure}
a {\em rank function} 
\index{matroid!rank function}
or a {\em girth function}.
\index{matroid!girth function}
All these definitions of matroids are equivalent from an axiomatic point of view, but
differ from an algorithmic point of view, cf. \cite{ar:HausmannKorte81,mayhew2008matroid}.

For convenience we repeat here also the definition of a matroid via a rank function from Chapter 8.
\ifmargin
\marginpar{\color{red} Chapter cross reference}
\else\fi 
{\color{red}
Chapter 8 (J. Oxley)
}
Let $E$ a finite set. A function $r: 2^E \rightarrow \Z$ is a {\em rank function}  if it satisfies
the following conditions:
\begin{description}
\item[(R1)]
If $X \subseteq E$ then $0 \leq r(X) \leq |X|$.
\item[(R2)]
If $X \subseteq Y \subseteq E$, then $r(X) \leq r(Y)$.
\item[(R3)]
If $X, Y \subseteq E$, then
$$
r(X) + r(Y) \leq r(X \cup Y) + r(X \cap Y)
$$
\end{description}
For $A \subseteq E$, 
$|A| - r(A)$ is called the {\em corank} or {\em nullity} of $A$.
A nonempty set $A \subseteq E$ is a circuit if  $r(A)= |A|+1$
and for every $e \in E$ we have that $r(A-\{e\}) = r(A)$.

Here is another equivalent definition of matroids using a girth function, 
cf. \cite{ar:HausmannKorte81}.
It is introduced here, because we will discuss its computational power as a matroid oracle in Theorem \ref{th:HausmannKorte}.
Let $E$ a finite set. A function $g: 2^E \rightarrow \N \cup \{\infty\}$ is a {\em girth function}  if it satisfies
the following conditions for all $X,Y \subseteq E$ and $x \in E$:
\begin{description}
\item[(G1)]
If $g(X) < \infty$ there is $Y \subseteq X$ with $g(X)=g(Y) = |Y|$.
\item[(G2)]
If $X \subseteq Y \subseteq E$, then $g(X) \geq g(Y)$.
\item[(G3)]
If $g(X) = |X|$, $g(Y) = |Y|$, $X \neq Y$ and $x \in X \cap Y$, 
then  \\ $g(X \cup Y - \{x\}) < \infty$.
\end{description}

In all the definitions of matroids
there is an exponential gap between the size of the underlying set $E$, and the information needed to
capture the properties of the family of subsets, the closure operation, the rank function or the girth function
defining the matroid.
In order to develop a complexity theory for matroid problems three approaches are used:
matroid {\em oracle computations} 
\index{matroid!oracle computations}
the restriction to matroids with {\em succinct presentations}, and using the full listing
of the family of subsets (circuits, independent sets, etc.) or the functions on the subsets
of the ground sets (rank function, closure operation, etc.).
Pioneering papers for the first two approaches are
\cite{ar:edmonds1965,ar:edmonds1971,bollobas1978packings,ar:RobinsonWelsh80,ar:HausmannKorte81,jensen1982complexity}.
More recent discussions can be found in \cite{mayhew2008matroid,snook2013matroids}.
\subsection{Succinct presentation of matroids}
Let $\mathcal{A}$ be a class of matroids closed under matroid isomorphisms, 
and $\Sigma^*$ be the finite words
over a finite alphabet $\Sigma$. Let $\ell(w)$ denote the length of $w \in \Sigma^*$.
and $\mathcal{A}_n$ the class of matroids in $\mathcal{A}$ with ground set 
$\{1, \ldots, n\}=[n]$
of size $n$.
A class of matroids has a 
{\em succinct description}, 
or is {\em succinct},  \index{matroid!succinct}
if there is an injective mapping
$e: \mathcal{A} \rightarrow \Sigma^*$ 
and a polynomial $p \in \mathbb{N}[X]$
such that for each $M \in \mathcal{A}_n$ we have that $\ell(e(M)) \leq p(n)$.
It is easy to see that $\mathcal{A}$ is succinct iff $|\mathcal{A}_n|= O(2^{q(n)})$ for some $q \in \mathbb{N}[X]$.

\begin{examples}
\label{CRC:succinct}
\ \\
\begin{enumerate}[(i)]
\item
The 
graphic matroids  \index{matroid!graphic}
are succinct using the underlying graph as their description.
\item
The binary matroids, \index{matroid!binary}
are succinct.
The same is true for the matroids which can be represented over any finite field.
\item
The 
regular matroids \index{matroid!regular}
are the matroids which are representable over every field.
Hence they form a succinct class of matroids, cf. {\color{red} Chapter 8, Theorem 8.16}.
If $M$ is regular, so is its dual.
Graphical matroids are regular.
\item
Transversal matroids, 
\index{matroid!transversal}
{\color{red} defined in Oxley's Chapter 8},
are representable over sufficiently large (finite) fields, hence they form
a succinct class of matroids (cf. \cite[Theorem 2.5]{bonin2010introduction}. 
Special cases of transversal matroids are
\begin{enumerate}
\item
The class of bicircular matroids \index{matroid!bicircular}. 
Given a graph $G$, its {\em bicircular matroid $B(G)$} is given by $E=E(G)$ and its independent sets
are the edge sets in which each connected component contains at most one cycle.
\item
The class of lattice-path matroids \index{matroid!lattice-path}.
A {\em lattice path matroid} is a transversal matroid that
has a presentation by a collection
of  intervals  in  the  ground  set,  relative  to  some  fixed  linear  order,  such  that  no  interval
contains another, cf. \cite[Section 4.2]{bonin2010introduction}.
\end{enumerate}
\ifmargin
\marginpar{\color{red} Chapter cross reference}
\else\fi 
\end{enumerate}
\end{examples}
In this chapter complexity results are always formulated for succinct classes of matroids.

The {\em uniform matroid} $U_{r,n}$ 
\index{matroid!uniform}
has an underlying set $E$ with $n$ elements and its basis consists of all the subsets
of $E$ with exactly $r$ elements. The uniform matroids are succinct.
We assume the reader is familiar with the following operations on matroids, cf.
\ifmargin
\marginpar{\color{red} Chapter cross reference}
\else\fi 
{\color{red}
Chapter 8 (J. Oxley)
}
Let $M,N$ be two matroids and $e,f$ be an edge of $M$ and $N$ respectively.
\begin{enumerate}[(i)]
\item
$M_{-e}$ is the matroid obtained from $M$ by {\em deleting} $e$.
\item
$M_{\backslash e}$ is the matroid obtained from $M$ by {\em contracting} $e$.
\item
$M \sqcup_{e,f} N$ is the $2$-sum of $M$ and $N$.
\item
$M \otimes N$ is the tensor product of $M$ and $N$.
Here $N$ is a {\em pointed matroid} with distinguished edge $d$.
In the case $N$ is a uniform matroid $U_{r,n}$ the result is independent of the choice of $d$
in $U_{r,n}$.
\end{enumerate}
Let $M$ be an arbitrary matroid. 
The {\em $k$-stretch $s_k(M)$} is defined as $s_k(M) = M \otimes U_{k,k+1}$,
and the {\em $k$-thickening $t_k(M)$} is defined as $s_k(M) = M \otimes U_{1,k+1}$.

\begin{definition}[\cite{ar:JVW90,vertigan1998bicycle}]
\label{CRC:JVW-class}
\begin{enumerate}[(i)]
\item
A class of matroids $\mathcal{C}$ is {\em closed under expansions}
\index{matroid class!closed under expansions}
if it is closed under $s_k(M)$ and $t_s(M)$ for every $k \in \N$.
\item
$\mathcal{C}$ is {\em expand-succinct} or a {\em $\JVW$-class}  if it is succinct and
$s_k(M)$ and $t_k(M)$ can be constructed in polynomial time for each $M \in \mathcal{C}$.
\end{enumerate}
\end{definition}

Typical $\JVW$ classes are the graphic matroids, the regular matroids,
and matroids representable over a fixed finite field. 
The class of transversal matroids is not a $\JVW$-class.
It is closed under deletion, but not under contraction, cf. Section 8.7 in Chapter 8.
\subsection{Matroid oracles}
Matroid oracles are used to show that most matroid properties are hard to compute for general matroids, cf. 
\cite{ar:RobinsonWelsh80,jensen1982complexity}. 
This is largely due to the fact that the number of matroids with underlying set $E$ of size $n$ is doubly exponential
in $n$.
Oracles may vary, cf. \cite{ar:HausmannKorte81}, but the most widely used 
is the {\em independence oracle},
\index{matroid!independence oracle}
which takes as its input a set of matroid elements, and returns as output a Boolean value, 
true if the given set is independent and false otherwise.
Naturally, every one of the nine axiomatic definitions of matroids,
$\mathrm{INDEPENDENT}$,
$\mathrm{BASIS}$,
$\mathrm{CIRCUIT}$,
$\mathrm{SPANNING}$,
$\mathrm{FLAT}$,
$\mathrm{HYPERPLANE}$,
$\mathrm{RANK}$,
$\mathrm{GIRTH}$ and
$\mathrm{CLOSURE}$,
gives rise to
an oracle. The oracles return Boolean values also for basis sets, circuits, 
spanning sets, flats and hyperplanes. In the case of the rank or girth function, the oracle
returns an element of $\{0, 1, \ldots, |E|, \infty\}$, and in the case
of the closure a subset of $E$.
The computational strength of the different oracles can be compared by polynomial time
simulation: Let $O_1, O_2$ be two oracles of matroids on a ground set $E$.
We say that $O_1$ is {\em polynomially reducible to} $O_2$, denoted by $O_1 \rightarrow O_2$,
iff one call in $O_1$ can be simulated by at most polynomially many calls on $O_2$.

\begin{theorem}[D. Hausmann and B. Korte, \cite{ar:HausmannKorte81}]
\label{th:HausmannKorte}
Under the partial order of polynomial reducibility we have:
\begin{enumerate}[(i)]
\item
$\mathrm{BASIS}$,
$\mathrm{CIRCUIT}$,
$\mathrm{FLAT}$,
$\mathrm{HYPERPLANE}$ are incomparable and weaker as oracles than
$\mathrm{RANK}$.
\item
$\mathrm{RANK}$,
$\mathrm{INDEPENDENT}$,
$\mathrm{SPANNING}$ and
$\mathrm{CLOSURE}$ polynomially bi-reducible.
\item
$\mathrm{GIRTH}$ is strictly stronger than 
all the other oracles.
\end{enumerate}
\end{theorem}

Many natural matroid parameters are hard to compute, \cite{jensen1982complexity}:
\begin{theorem}
\label{th:hard-matroids}
The following are not polynomial-time computable using the $\mathrm{INDEPENDENT}$-oracle:
\begin{enumerate}[(i)]
\item
Decide whether $M$ is uniform, self-dual, orientable, bipartite, Eulerian or representable.
\item
Compute the girth or the connectivity  of $M$.
\item
Count the number of circuits, bases, hyperplanes or flats of $M$.
\end{enumerate}
\end{theorem}
However, to the best of our knowledge, there are no papers studying the degrees  of computability
for oracle computations on matroids.

\subsection{Full presentations of matroids}
A matroid is essentially a finite set with a structured family of subsets, 
which can be described explicitly.
This approach has originally not been studied, because the input for a Turing machine
was considered too large, if 
the size of the matroid should be polynomial in the size of
the cardinality of the ground set $E$. 
It was suspected, that this would make most, if not all computational matroid problems
solvable in polynomial time in the size of its true input.
However, D. Mayhew analyzed the situation carefully, \cite{mayhew2008matroid},
by comparing the various input sizes of
$\mathrm{RANK}$,
$\mathrm{INDEPENDENT}$,
$\mathrm{SPANNING}$,
$\mathrm{BASIS}$,
$\mathrm{CIRCUIT}$,
$\mathrm{FLAT}$ and
$\mathrm{HYPERPLANE}$.

Suppose that 
$\mathrm{INPUT}_1$
and
$\mathrm{INPUT}_2$ are two methods for describing a matroid $M$.
Then
$\mathrm{INPUT}_1 \leq \mathrm{INPUT}_2$ iff there exists a polynomial-time Turing machine
which will produce  for every matroid $M$ the presentation
$(M,\mathrm{INPUT}_2)$ given the presentation 
$(M,\mathrm{INPUT}_1)$.
We write
$\mathrm{INPUT}_1 < \mathrm{INPUT}_2$ iff it is not the case
that
$\mathrm{INPUT}_2 \leq \mathrm{INPUT}_1$. 
Surprisingly, comparing input modes gives a different picture than comparing
matroid oracles.

\begin{theorem}[D. Mayhem, \cite{mayhew2008matroid}]
Under the partial ordering of $\leq$ we have
\begin{enumerate}[(i)]
\item
$\mathrm{RANK} < \mathrm{INDEPENDENT}$,
and
$\mathrm{RANK} < \mathrm{SPANNING}$.
Furthermore
$\mathrm{INDEPENDENT}$ and
$\mathrm{SPANNING}$ are incomparable.
\item
$\mathrm{SPANNING} < \mathrm{BASIS} < \mathrm{CIRCUIT}$ and
$\mathrm{BASIS} < \mathrm{HYPERPLANES}$. 
Furthermore,
$\mathrm{CIRCUITS}$ and $\mathrm{HYPERPLANES}$ are incomparable.
\item
$\mathrm{INDEPENDENT} < \mathrm{FLATS} < \mathrm{HYPERPLANES}$ and
$\mathrm{INDEPENDENT} < \mathrm{BASIS}$. 
Furthermore,
$\mathrm{BASIS}$ and $\mathrm{FLATS}$ are incomparable.
\end{enumerate}
\end{theorem}

\subsection{Comparing notions of matroid computability}
For succinct presentations of matroids decision and counting problems
can be found in most time-complexity classes between polynomial time
and exponential time, because this includes graph matroids.

Complexity results for matroid oracle are usually of the form
\begin{quote}
There exists no polynomial $\mathrm{ORACLE}$-algorithm for computing $\mathrm{PROBLEM}$.
\end{quote}
In \cite{ar:HausmannKorte81} twenty problems are listed for which this is true 
for $\mathrm{INDEPENDENT}$, among them also the problem of computing the
Tutte polynomial for arbitrary matroids.

It remains open how to work out the details needed to refine complexity results
for matroid oracles analogue to the case of succinct presentations.
For which problems is there an analogue to $\bNP$-completeness or
$\sP$-completeness in the oracle setting.

Finally, in the case of full descriptions of the matroids, it is not the case that,
as one might naively suspect, that all natural problems are in $\bP$.
The situation is well illustrated by the following problem:

\begin{quote}
{\bf Problem:} 3-matroid-intersection $\mathrm{3MI}$
\\
{\bf Instance:} An integer $k$ and three matroids $(M_i, \mathrm{INPUT})$, $1 \leq i \leq 3$
over the same ground set $E$.
\\
{\bf Question:} Does there exist a set $A \subseteq E$ such that $|A|=k$ and $A$
is independent in each $M_i$?
\end{quote}

\begin{theorem}[D. Mayhem, \cite{mayhew2008matroid}]
\ \\
\begin{enumerate}[(i)]
\item
If 
$\mathrm{CIRCUITS} \leq \mathrm{INPUT}$ or
$\mathrm{HYPERPLANES} \leq \mathrm{INPUT}$, then  
$\mathrm{3MI}$
is $\bNP$-complete.
\item
If
$\mathrm{INPUT} \leq \mathrm{BASIS}$, then
$\mathrm{3MI}$
is in $\bP$.
\end{enumerate}
\end{theorem}

To the best of our knowledge, the complexity of the Tutte polynomial
has not been established in the full description framework.

\subsection{Matroid width}
\label{KM:se:matroid-width}

J. Oxley and D. Welsh in \cite{oxley1992tutte} introduce a very restriced notion of matroid width:
A class of matroids $\mathcal{M}$  has OW-width $k$ if a largest $3$-connected member of 
$\mathcal{M}$ has a groundset of $k$ elements.
\index{matroid!OW-width}
They show that, under some stronger assumption than succinct representability,
evaluating the Tutte polynomial on matroid classes of bounded OW-width a in $\FPT$.
We shall not persue this further.

A {\em branch decomposition}
\index{matroid!branch decomposition}
of a matroid $M$ with ground set $E$ and rank function $r$ is a tree $T$ such that
\begin{enumerate}[(i)]
\item
all inner nodes of $T$ have degree three, and
\item
the leaves of $T$ are in one-one correspondence with $E$.
\end{enumerate}
An edge $e$ of $T$ splits $T$ into two subtrees with leaves corresponding to  
$E_1(e)$ and $E_2(e)$ which partition $E$. The width of $e$ is defined by 
$r(E_1(e))+ r(E_2(e))- r(E)+ 1$. 
The {\em width of the branch decomposition} $T$ is the maximum width of an edge of $T$. 
Finally the {\em branch-width of} $M$ is the minimum width of the
branch decompositions of $M$, and is denoted by $\mathrm{bw}(M)$.
\index{matroid!branch-width}

We collect some basic facts listed in the survey paper \cite{ar:HlinenyOumSeeseGottlob08}. 

\begin{theorem}
\begin{enumerate}[(i)]
\item
The branch-width of a bridgeless graph (every edge is contained in a cycle) equals the
branch-width of its cycle matroid.
\item
Given a matroid $M$ with $n$ elements, and a positive integer $k$, 
it is possible to test,
using an $\mathrm{INDEPENDENT}$-oracle, 
in polynomial-time (with the degree depending on $k$), 
whether $\mathrm{bw}(M) \leq k$.
\item
For  matroids $M$ representable over a fixed finite field, deciding 
whether $\mathrm{bw}(M) \leq k$ is in $\bFPT$ (fixed parameter tractable).
If so, then the same algorithm also outputs a branch-decomposition of $M$ of width
at most $3k$.
\end{enumerate}
\end{theorem}
Although testing whether $\mathrm{bw}(M) \leq k$ is easy to check in the oracle framework,
applications usually require that the matroid be representable, which is hard 
by Theorem \ref{th:hard-matroids}.
To remedy this, other notions of matroid width were recently introduced:
{\em decomposition width} in \cite{kral2012decomposition} 
\index{matroid!decomposition width}
and {\em amalgam width} in \cite{mach2013amalgam}.
\index{matroid!amalgam width}
Without going into details we note that matroids of decomposition width  $\mathrm{dw}(M) =k$
(amalgam width $\mathrm{aw}(M) =k$)
can be represeted by a {\em decomposition tree} ({\em amalgam tree}), although this tree  is not
known to be computable in polynomial time.
However, if the decomposition or amalgam tree is given together with $M$ and is of width $k$, 
then many parameters can be computed in polynomial time using the 
$\mathrm{INDEPENDENT}$-oracle, \cite{kral2012decomposition,mach2013amalgam}.
We also note that for matroids $M$  representable by a finite field $\mathbb{F}$ 
and branch-width $\mathrm{bw}(M) \leq k$
both $\mathrm{dw}(M)$ and $\mathrm{aw}(M)$ are bounded by functions which depend on $k$
and the size of $\mathbb{F}$.

\ifmargin
\marginpar{\color{red} file: ARXIV-tmatroids.tex}
\else\fi 
\section{The exact complexity of the Tutte polynomial for matroids}
\label{KM:se:tmatroids}
The paper \cite{ar:JVW90} was the first to analyze the complexity of evaluating
the Tutte polynomial for pairs of algebraic numbers $x_0,y_0 \in \mathbf{A}$ also for matroids.
The Tutte polynomial of a graphic matroid $M(G)$ is the same as the Tutte polynomial of $G$.
So here we only discuss classes of non-graphic matroids.

\subsection{Succinct presentations}
For the case of a class $\mathcal{C}$ matroids, the analogue of Theorem \ref{CRC:JVW-theorem} 
uses two assumptions on the class $\mathcal{C}$ which are trivially satisfied for the class
$\mathcal{G}$ of all graphic matroids.
We have introduced these assumptions 
in Section \ref{KM:se:matroids} to define a $\JVW$-class of matroids.
A $\JVW$-class has to be succinct, closed under $k$-stretching and $k$-thickening for all $k \in \N$,
the operations of stretching and thickening have to be computable
in polynomial time.

\begin{theorem}[F. Jaeger, D. Vertigan and D. Welsh \cite{ar:JVW90}] 
\label{CRC:JVW-theorem-m}
Let $\mathcal{C}$ be $\JVW$-class of matroids and $M \in \mathcal{C}$. 
Then evaluating $T(M;a,b)$ is
$\sP$-hard on all points $(a,b) \not \in H$. 
If $(a,b)\in H$, then $T(M;a,b)$ is computable in polynomial time.
\end{theorem}
Here $H$ is the same as in Theorem \ref{CRC:JVW-theorem}.
\index{matroid!JVW-class}

The class of transversal matroids $\mathcal{T}$ is succinct, cf. Example \ref{CRC:succinct}.
However, it is not a $\JVW$-class. We still have the following dichotomy:

\begin{theorem}[C. Colbourn, J. Provan and D. Vertigan, \cite{colbourn1995complexity}]
For the class of transversal matroids $M \in \mathcal{T}$ the evaluation the Tutte polynomial
$T(M,x_0,y_0)$ is $\sP$-hard for all $(x_0,y_0) \in \mathbf{A}$, unless $(x_0-1)(y_0-1)=0$,
in which case it is in $\bP$.
\end{theorem}
\index{matroid!transversal}

To prove this theorem, the authors replace the operations $k$-stretching and $k$-thickening
used in the proof of  Theorem \ref{CRC:JVW-theorem-m} by operations $k$-expansion and $k$-augmentation
which preserve transversality.

A {\em bicircular matroid} 
\index{matroid!bicircular}
of a graph $G$ is the matroid $B(G)$ whose points are the edges of 
$G$ and whose independent sets are the edge sets of pseudoforests of $G$, 
that is, the edge sets in which each connected component contains at most one cycle.
The class $\mathcal{B}$ of bicircular matroids is contained in the class $\mathcal{T}$ 
of transversal matroids, which, unlike transversal matroids,
is closed under minors. A bicircular matroid $B =B(G)$ is determined by its underlying graph $G$,
but it need not be  graphic. The $k$-stretching of a bicircular matroid $B$, denoted by $s_k(B)$
can be shown to be $B(s_k(G))$. Hence, bicircular matroids are closed under $k$-stretching.
However, they are not closed under $k$-thickening.

\begin{theorem}[\cite{gimenez2006complexity}]
For a matroid
$M \in \mathcal{B}$, the evaluation of the Tutte polynomial
$T(M,X,Y)$ is $\sP$-hard for all $(x_0,y_0) \in \mathbf{A}$, unless $(x_0-1)(y_0-1)=0$.
For
$x_0=0$ and $x_0 = -1$ the complexity has not yet been determined.
\end{theorem}

\begin{proposition}[\cite{gimenez2006complexity}]
For the complete graph $K_n$ and the bicircular matroid $M=B(K_n) \in \mathcal{B}$,
the evaluation of the Tutte polynomial
$T(M,x_0,y_0)$ is  in $\P$.
\end{proposition}

A further special case of transversal matroids is the class of the {\em lattice-path matroids} 
$\mathcal{LP}$.
\index{matroid!lattice path}
They are obtained from the lattice with vertices in $(i,j) \in\N^2$  with 
$0 \leq i \leq m$ and
$0 \leq j \leq r$ with steps $\mathrm{EAST}$ and $\mathrm{NORTH}$.
Let $P$ and $Q$ be two paths in this lattice. 
The lattice paths that go from $(0,0)$ to $(m, r)$ and that remain in the region bounded by 
$P$ and $Q$ can be identified with the bases of a particular type of transversal matroid, 
\cite{bonin2006lattice}.

\begin{theorem}[\cite{morton2013computing}]
For lattice-path matroids
$M \in \mathcal{LP}$, the evaluation of the Tutte polynomial
$T(M,x_0,y_0)$ is  in $\P$.
\end{theorem}

\subsection{Matroid oracles}

The situation for matroids which have no succinct presentation
the situation is less understood.

We summarize here what is known about the computability of the Tutte polynomial
when the matroid is given by an oracle.

\begin{theorem}
\label{th:oraclematroid}
Let $M$ be a matroid of size $n$ and rank $r$. 
Using the $\mathrm{INDEPENDENT}$-oracle we have
\begin{enumerate}[(i)]
\item
The Tutte polynomial $M$ is not computable in polynomial time, \cite{jensen1982complexity}.
\item
If the the matroid $M$ has branch-width $\mathrm{bw}(M) \leq k$ and is representable over
a finite field $\mathbb{F}$, then $T(M,x,y)$ is computable in time
$O(n^6 \log(n) \log\log(n))$ even without the oracle, \cite{hlineny2006tutte}.
\item
If $M$ has decomposition-width $\mathrm{dw}(M) \leq k$ and
is given with its decomposition tree, then $T(M,x,y)$ is computable in
time $O(k^2n^3r^2)$ and evaluated in time $O(k^2n)$ (under the assumption of unit cost
of the arithmetic operations), \cite{kral2012decomposition}.
\item
If $M$ has amalgam-width $\mathrm{aw}(M) \leq k$ and
is given with its amalgam tree, then $T(M,x,y)$ is computable in
time  $O(n^d)$ for some constant $d$ independent of $k$, \cite{mach2013amalgam}.
\end{enumerate}
\end{theorem}

\index{computation model!Turing|)}
\ifmargin
\marginpar{\color{red} file: FINAL-algebraic.tex}
\else\fi 
\section{Tutte polynomial in algebraic models of computation}
The complexity of the Tutte polynomial was also studied in various
alternative models of computation.

\subsection{Valiant's algebraic circuit model}
\index{computation model!algebraic}
L. Valiant introduced in \cite{Val} an algebraic model of computation based on uniformly
defined families of algebraic circuits and a notion of reducibility 
based on substitutions called {\em  p-projections}.
In this framework it is natural to consider multivariate versions of graph polynomials,
see also
{\color{red} see also Chapter 18 (Traldi)}.
In this model there is analogue for polynomial time called $\VP$ 
\index{complexity!VP}
and for non-deterministic polynomial
time called $\VNP$. 
\index{complexity!VNP}
Computing the family of determinants $\mathrm{DET}_n$ of an $(n \times n)$ matrix where the entries
are treated as indeterminates is in $\VP$,
\index{determinant}
computing the corresponding permanent $\mathrm{PER}_n$ is $\VNP$-complete.
\index{permanent}
The major problem in Valiant's framework is the question whether $\VP = \VNP$.
A detailed development of Valiant's theory can be found in \cite{bk:Buergisser00}.

Typical families of functions in $\VNP$ are generating functions of graph properties.
These are polynomials depending on graphs with indeterminate weights on the edges.
Let $\mathcal{P}$ be a graph property, i.e., a class of finite graphs closed under isomorphisms.
For a graph $G=(V(G), E(G))$
the generating function of a graph property is defined as
$$
GF(G,\mathcal{P}) =
\sum_{E' \subseteq E, (V(G), E') \in \mathcal{P}}
\prod_{e \in E'} X_e
$$
The complexity of generating functions of graph properties in Valiant's model was first studied by M. Jerrum
in \cite{jerr:81} and further developed by P. B\"urgisser in \cite{bk:Buergisser00}.

\index{weighted Tutte polynomial}
The Tutte polynomial of a graph has only two variables, therefore it cannot be $\VNP$-complete in Valiant's model.
In order to get a complete family of Tutte polynomials we have to put weights on the edges of the
underlying graph. This leads to a weighted Tutte polynomial as defined in \cite{Traldi89} or in \cite{bollobas-riordan1999tutte}.
Let $G=(V(G),E(G))$ be a graph. For each $e \in E(G)$ we have four indeterminates $X_e, Y_e, x_e, y_e$.
Fix an order on the edges $E(G)$.
The multivariate 
Whitney rank generating function
$W(G_n; X_e,Y_e, x_e, y_e)$ is defined for connected graphs $G$ by
\begin{gather}
W(G_n; X_e,Y_e, x_e, y_e)= \notag \\
\sum_{T \subseteq E(G)}
\prod_{e \in T, int-act} X_e
\prod_{e \in E(G)-T, ext-act} Y_e
\prod_{e \in T, int-inact} x_e
\prod_{e \in E(G)-T, ext-inact} y_e
\notag
\end{gather}
where the sum is over all spanning trees $T$, 
and the condition of internal/external active/inactive is defined with respect to the order on $E(G)$.
For graphs not connected we add additional indeterminates.

\index{Traldi's dichromatic polynomial}
Traldi's dichromatic  polynomial  $Q(G;v_e,q,z)$ is defined by
$$
Q(G;v_e,q,z) =
\sum_{E' \subseteq E(G)} q^{k(E')}z^{|E'| - r(E')} \prod_{e \in E'} v_e
$$
where $k(E')$ is the number of connected components of the spanning subgraph induced by $E'$,
and $r(E')$ is its rank.

The complexity of the Tutte polynomial in Valiant's model was studied in \cite{lotz2004algebraic}.
To get a meaningful complexity analysis, the notion of p-projections is modified to allow also
a wider class of substitutions which the authors call {\em polynomial oracle reductions}.
Not surprisingly, they show that in this (slightly) modified framework of Valiant's model the
multivariate Tutte polynomial is $\VNP$-complete. More precisely:

\index{complexity!VNP-completeness}
\begin{theorem}[\cite{lotz2004algebraic}]
\begin{enumerate}
\item
There is a family of graphs $G_n$ such that the family of multivariate Tutte polynomials
$W(G_n; X_e,Y_e, x_e, y_e)$ is $\VNP$-complete under p-projections.
\item
Traldi's dichromatic  polynomial 
$ Q(G;v_e,q,z) $
is $\VNP$-complete via polynomial oracle reductions.
\end{enumerate}
\end{theorem}
For details the reader is referred to \cite{lotz2004algebraic,phd:Hoffmann}.
\subsection{The Blum-Shub-Smale model}
\index{computation model!Blum-Shub-Smale (BSS)}
In the early 1970s  computability over arbitrary first order structures was introduced independently
in \cite{engeler1967algorithmic,friedman1970algorithmic} by E. Engeler and H. Friedman. 
What they propose is, roughly speaking, to use register machines over first order structures $\mathfrak{A}$,
where the registers contain elements of $\mathfrak{A}$ and the tests and operations are defined by the
relations and functions of $\mathfrak{A}$.
Their framework was rediscovered by L. Blum, M. Shub and S. Smale in \cite{blss:89}, see also \cite{bk:BCSS} and
\cite{friedman1992algorithmic} for a comparison between \cite{friedman1970algorithmic} and \cite{blss:89}.
They main merit of \cite{blss:89} 
is the introduction of
a complexity theory for computability
over the real and complex numbers which includes complexity classes 
$\bP_{\mathcal{R}}$ for deterministic polynomial-time computable and
$\bNP_{\mathcal{R}}$ for non-deterministic polynomial-time computable problems, and the existence of
$\bNP_{\mathcal{R}}$-complete problems, where $\mathcal{R}$ is an arbitrary 
(possibly ordered) ring.
We call this model of computation the {\em $\mathrm{BSS}$ model of computation}.
\index{computation model!Blum-Shub-Smale (BSS)|(}
Later, K. Meer, \cite{ar:Meer00}, introduced also a complexity class $\sharp\bP_{\mathcal{R}}$ for 
the $\mathrm{BSS}$ model of computation.

As it is customary to look at the Tutte polynomial $T(G;x,y)$ as a polynomial in 
$\R[x,y]$ or
$\C[x,y]$, and graphs can be viewed as $0-1$-matrices, the complexity of the Tutte polynomial
is most naturally discussed in the $\mathrm{BSS}$ model of computation,
\cite{pr:MakowskyMeer2000}. 
However, the theory of counting complexity in the $\mathrm{BSS}$ model of computation 
has various drawbacks. 
In particular, there is no satisfactory theory for $\sharp\bP_{\mathcal{R}}$-complete
problems. Furthermore, it is not clear whether, and it seems unlikely that, 
evaluating the chromatic polynomial at the point $x=3$
is $\bNP_{\mathcal{R}}$-complete.
A detailed description  of these problems can be found in
\cite{ar:KotekMakowskyRavve2012,pr:KotekMakowskyRavve2013}, 
where also an abstract algebraic version of the JVW-Theorem, \ref{CRC:JVW-theorem},
is formulated. In
\cite{arenas2017descriptive} 
the descriptive complexity of counting problems in the $\mathrm{BSS}$ model of computation over $\R$ 
is also discussed using essentially the same framework as in 
\cite{ar:KotekMakowskyRavve2012,pr:KotekMakowskyRavve2013}. 
\index{computation model!Blum-Shub-Smale (BSS)|)}


\ifmargin
\marginpar{\color{red} file: FINAL-problems.tex}
\else\fi 
\section{Open problems}
The complexity of computing or evaluating the Tutte polynomial 
in the Turing model of computation
in a ring $\mathcal{R}[x,y]$
is well understood
\begin{itemize}
\item
for graphs without restrictions on their presentations, and
\item
for matroids, provided the matroids have succinct presentations,
\end{itemize}
provided the arithmetic operations in the ring $\mathcal{R}$ 
are polynomial-time computable in some standard presentation of $\mathcal{R}$.
The same is true for special classes of graphs (planar, bipartite, etc.) 
and special succinct classes of matroids (transversal, bicircular, etc.).
For graph classes of bounded tree-width evaluating the Tutte polynomial is fixed parameter
tractable. However, for graph classes of bounded clique-width, the situation is not completely understood.
\begin{problem}
Determine the complexity of evaluating the Tutte polynomial for graphs of fixed clique-width.
\end{problem}

Planar graphs are a special cases of a minor-closed class of graphs.
\begin{problem}
Determine the complexity of evaluating the Tutte polynomial for other minor-closed graph classes.
\end{problem}

The JVW Theorem, \ref{CRC:JVW-theorem}, and its variations show a 
dichotomy, the {\em Difficult Point Property ($\mathrm{DPP}$)}: 
\index{Difficult Point Property}
Evaluation of the Tutte polynomial
at fixed evaluation points is $\sharp\bP$ hard on all points in $\mathcal{R}^2 - A$ where
$A \subset \mathcal{R}^2$ is
a semialgebraic (quasialgebraic) set of dimension $1$.
Versions of $\mathrm{DPP}$ have been proven also for
many other graph polynomials, e.g., the Bollob\'as-Riordan version of the Tutte polynomial,
\cite{ar:BlaeserDellMakowsky08,ar:BlaeserDellMakowsky10},
the interlace polynomial,
\cite{ar:BlaeserHoffmann07,pr:BlaserHoffmann08},
the cover polynomial for directed graphs,
 \cite{pr:BlaserDell07,ar:BlaeserDellFouz2012}.
the bivariate matching polynomial
for multi-graphs defined first in \cite{ar:HeilmannLieb72} in \cite{ar:AverbouchMakowsky07,ar:Hoffmann10},
and many more which are surveyed in 
\cite{pr:KotekMakowskyRavve2013}.

\begin{problem}
Characterize the graph polynomials for which $\mathrm{DPP}$ holds.
\end{problem}

If one studies the complexity in the Turing model of computation, the real or complex numbers
have no recursive presentation. Moreover, the recursive presentaions of the 
countable subrings which do have a recursive presentation are not uniform. This makes
the statement about the complexity over the reals or complex numbers somehow artificial.
In the $\mathrm{BSS}$ model of computation, cf. \cite{pr:KotekMakowskyRavve2013},
this problem of non-uniformity does not arise.

\begin{problem}
Refine the complexity analysis of the Tutte polynomial in the $\mathrm{BSS}$ model of computation.
\end{problem}

In the case of matroids without succinct presentation, it is known that the Tutte polynomial
is not polynomial-time computable using the 
the $\mathrm{INDEPENDENT}$-oracle, Theorem \ref{th:oraclematroid}.
But to the best of our knowledge, 
the analogues of $\bNP$-completeness, or higher levels of the complexity
hierarchy within exponential time oracle computability of matroids has not been developed.

\begin{problem}
Develop a coherent theory of complexity for  computations with matroid oracles in general.
\end{problem}


\bibliographystyle{plain}
\bibliography{FINAL}
\printindex
\end{document}